\documentstyle[12pt,epsfig]{article}

\newskip\humongous \humongous=0pt plus 1000pt minus 1000pt

\newif\ifdtup

\def\abs#1{\left| #1\right|}

\def\beq{\begin{equation}}
\def\eeq{\end{equation}}

\def\beqn{\begin{eqnarray}}
\def\eeqn{\end{eqnarray}}
\relax
\def\bt{\tilde\beta}
\def\b{\beta}

\def\dotx{\dotx{\dot\overline{x}}}

\relax

\catcode`\@=11

\@addtoreset{equation}{section}
\def\theequation{\thesection\arabic{equation}}

\def\@normalsize{\@setsize\normalsize{15pt}\xiipt\@xiipt
\abovedisplayskip 14pt plus3pt minus3pt%
\belowdisplayskip \abovedisplayskip
\abovedisplayshortskip \z@ plus3pt%
\belowdisplayshortskip 7pt plus3.5pt minus0pt}

\def\small{\@setsize\small{13.6pt}\xipt\@xipt
\abovedisplayskip 13pt plus3pt minus3pt%
\belowdisplayskip \abovedisplayskip
\abovedisplayshortskip \z@ plus3pt%
\belowdisplayshortskip 7pt plus3.5pt minus0pt
\def\@listi{\parsep 4.5pt plus 2pt minus 1pt
     \itemsep \parsep
     \topsep 9pt plus 3pt minus 3pt}}

\@twosidetrue

\relax

\catcode`@=12

\catcode`\@=11

\def\section{\@startsection{section}{1}{\z@}{3.5ex plus 1ex minus
   .2ex}{2.3ex plus .2ex}{\large\bf}}

\def\thesection{\arabic{section}.}

\def\appendix{\setcounter{section}{0}
 \def\thesection{APPENDIX \Alph{section}:}
 \def\theequation{\Alph{section}.\arabic{equation}}}

\def\ps@headings{\def\@oddfoot{}\def\@evenfoot{}
\def\@oddhead{\hbox{}\hfill
 \makebox[.5\textwidth]{\raggedright\ignorespaces --\thepage{}--
 \hfill {}}}  
\def\@evenhead{\@oddhead}
\def\subsectionmark##1{\markboth{##1}{}}
}

\ps@headings

\catcode`\@=12

\def\figcap{\section*{Figure Captions\markboth
 {FIGURECAPTIONS}{FIGURECAPTIONS}}\list
 {Fig. \arabic{enumi}:\hfill}{\settowidth\labelwidth{Fig. 999:}
 \leftmargin\labelwidth
 \advance\leftmargin\labelsep\usecounter{enumi}}}
 \relax
\def\tablecap{\section*{Table Captions\markboth
 {TABLECAPTIONS}{TABLECAPTIONS}}\list
 {Table \arabic{enumi}:\hfill}{\settowidth\labelwidth{Table 999:}
 \leftmargin\labelwidth
 \advance\leftmargin\labelsep\usecounter{enumi}}}
 \relax
\def\reflist{\section*{References\markboth
 {REFLIST}{REFLIST}}\list
 {[\arabic{enumi}]\hfill}{\settowidth\labelwidth{[999]}
 \leftmargin\labelwidth
 \advance\leftmargin\labelsep\usecounter{enumi}}}
 \relax

\catcode`\@=11

\def\ps@headings{\def\@oddfoot{}\def\@evenfoot{}
\def\@oddhead{\hbox{}\hfill
 \makebox[.5\textwidth]{\raggedright\ignorespaces --\thepage{}--
 \hfill {}}}    
\def\@evenhead{\@oddhead}
\def\subsectionmark##1{\markboth{##1}{}}
}

\ps@headings

\catcode`\@=12

\relax

\catcode`\@=11
\def\prm{\fam \z@}
\catcode`\@=12

\relax
\def\pl#1#2#3{{\it Phys. Lett. }{\bf #1}(19#2)#3}
\def\zp#1#2#3{{\it Z. Phys. }{\bf #1}(19#2)#3}

\def\np#1#2#3{{\it Nucl. Phys. }{\bf #1}(19#2)#3}

\def\aop#1#2#3{{\it Ann. Phys. }{\bf #1}(19#2)#3}

\def    \hepph  #1 {{\tt hep-ph/#1}}
\def    \hepex  #1 {{\tt hep-ex/#1}}

\relax

  \newcommand{\ccaption}[2]{
    \begin{center}
    \parbox{0.85\textwidth}{
      \caption[#1]{\small{\it{#2}}}
      }
    \end{center}
    }
\begin{document}
\def\theequation{\arabic{equation}}          
\newcommand\sss{\scriptscriptstyle}
\newcommand\me{m_e}
\newcommand\as{\alpha_{\sss S}}         
\newcommand\aem{\alpha_{\rm em}}
\newcommand\refq[1]{$^{[#1]}$}
\renewcommand\topfraction{1}       
\renewcommand\bottomfraction{1}    
\renewcommand\textfraction{0}      
\setcounter{topnumber}{5}          
\setcounter{bottomnumber}{5}       
\setcounter{totalnumber}{5}        
\setcounter{dbltopnumber}{2}       
\newsavebox\tmpfig
\newcommand\settmpfig[1]{\sbox{\tmpfig}{\mbox{\ref{#1}}}}
%
\begin{titlepage}
\nopagebreak
\vspace*{-1in}
{\leftskip 11cm
\normalsize
\noindent   
\newline
CERN-TH/97-255 \\
GEF-TH-9/1997 \\
hep-ph/9709481

}
\vskip 1.0cm
\vfill
\begin{center}
{\large \bf The gluon contribution}\\ 
{\large \bf to the polarized structure function $g_2$}
\vfill
\vskip .6cm 
{\bf Andrea Gabrieli}
\vskip .2cm
{\it Dipartimento di Fisica, Universit\`a di Genova,}\\
{\it Via Dodecaneso 33, I-16146 Genoa, Italy}
\vskip .5cm                                               
{\bf Giovanni Ridolfi}
\vskip .2cm
{\it CERN, TH Division, CH-1211 Geneva 23, Switzerland}\\
{\it and INFN Sezione di Genova, Via Dodecaneso 33, I-16146 Genoa, Italy}

\end{center}
\vfill
\nopagebreak
\begin{abstract}
{\small
We compute the structure function $g_2$ for a gluon target 
in perturbative QCD at order $\as$. We show that its first moment
vanishes, as predicted by the Burkhardt-Cottingham sum rule.
}
\end{abstract}
\vfill
CERN-TH/97-255 \\
September 1997    \hfill
\end{titlepage}

A computation of the structure function $g_2$ for a target quark
at order $\as$ in perturbative QCD has been performed in ref.~\cite{ALNR}.
The interest in that computation was mainly driven by the possibility
of performing a direct test of the validity of the Burkhardt-Cottingham (BC)
sum rule~\cite{BC} in perturbative QCD. The BC sum rule states that
\beq
\label{bcsr}
\int_0^1 dx\,g_2(x,Q^2)=0
\eeq
on the basis of general arguments about the analytic structure
in the complex $\nu\equiv Q^2/(2x)$ plane of the amplitude whose
imaginary part gives $g_2$
(a detailed discussion on the derivation and the validity of the BC sum
rule is given in ref.~\cite{Jaffe}).
It is therefore interesting to check whether eq.~(\ref{bcsr})
is valid in perturbative QCD.
The result of ref.~\cite{ALNR} (later confirmed in ref.~\cite{Kodaira2})
is that the first moment of $g_2$
actually vanishes at order $\as$ for a massive target quark
(at leading order, $g_2$ itself vanishes for a target quark).

In this note we present an analogous calculation for a gluon target.
We will show that also in this case the BC moment vanishes 
in a wide class of regularization schemes for the collinear singularities.

Structure functions are defined starting from the Fourier transform
of the forward matrix element of the product of two electromagnetic
currents between polarized states:
\beq
W^{\mu\nu}(p,q,s)=\frac{1}{4\pi}
\int d^4x\,e^{iqx}\,\langle p,s|J^\mu(x) J^\nu(0)|p,s\rangle,
\eeq
where $p$ and $s$ are the target momentum and spin four-vectors, respectively. 
In the case of a target gluon, the tensor $W^{\mu\nu}$ is given by
\beq
W^{\mu\nu}(p,q,s)=W^{\mu\nu\rho\sigma}(p,q)\epsilon_\rho\epsilon_\sigma^*,
\eeq
where the gluon polarization vector $\epsilon$ carries the dependence on $s$.
Only the antisymmetric part of $W^{\mu\nu}$ is relevant for the computation of
polarized structure functions; we have
\beq
iW_A^{\mu\nu}=\frac{1}{2}(W^{\mu\nu}-W^{\nu\mu})
=\frac{1}{2}(W^{\mu\nu\rho\sigma}-W^{\nu\mu\rho\sigma})
\epsilon_\rho\epsilon_\sigma^*
=W^{\mu\nu\rho\sigma}\frac{1}{2}(\epsilon_\rho\epsilon_\sigma^*
-\epsilon_\sigma\epsilon_\rho^*),
\eeq
where in the last step we used the symmetry of
$W^{\mu\nu\rho\sigma}$ under the simultaneous exchanges
$\mu\leftrightarrow\nu,\rho\leftrightarrow\sigma$.
We are therefore interested in computing the antisymmetric part
of the gluon polarization density matrix 
$\rho_{\rho\sigma}=\epsilon_\rho\epsilon^*_\sigma$.
To do this, we assume that the gluon in the initial state is off the
mass shell,
$p^2\ne 0$; this allows us to
define a longitudinal spin vector for the gluon:
\beq
s^\alpha = \lambda N\left(p^\alpha-\frac{p^2}{pq}\,q^\alpha\right),
\label{spin}
\eeq
which satisfies the transversity condition $ps=0$; $\lambda$ is the gluon
helicity, and the normalization factor $N$ (real and positive)
is related to $s^2$ by
\beq
N^2=-\frac{s^2}{p^2\bt^2},
\label{spinnorm}
\eeq
where
\beq
\bt=\sqrt{1-\frac{p^2 q^2}{(pq)^2}}=\sqrt{1+\frac{4p^2 x^2}{Q^2}},
\eeq
with the usual definitions $Q^2=-q^2$, $x=Q^2/(2pq)$.
We see that we can choose $N$ so that $\abs{s^2}=1$, provided $s^2$
and $p^2$ have opposite signs. We can now write
the antisymmetric component of $\epsilon_\rho\epsilon^*_\sigma$
as a function of $p$ and $s$:
\beq
\frac{1}{2}(\epsilon_\rho\epsilon^*_\sigma-\epsilon_\sigma\epsilon^*_\rho)
=-\frac{i}{2\sqrt{\abs{p^2}}}\,
\epsilon_{\rho\sigma\alpha\beta}p^\alpha s^\beta;
\eeq
the normalization is fixed by the condition
that, for $p^2\to 0^+$, 
the imaginary part of $\rho_{12}$ is equal to $-\lambda/2$,
where $\lambda$ is the gluon helicity.

The structure functions $g_1$ and $g_2$ are conventionally defined by means
of the following, general parametrization of $iW_A^{\mu\nu}$:
\beq
iW_A^{\mu\nu}=
\frac{i\sqrt{\abs{p^2}}}{pq}\epsilon^{\mu\nu\rho\sigma}q_\rho
\left[g_1(x,Q^2) s_\sigma
     +g_2(x,Q^2)\left(s_\sigma-\frac{qs}{pq}p_\sigma\right)\right],
\eeq
where $s$ is now a generic spin vector (not necessarily longitudinal,
since the gluon is off the mass shell).
One can define projectors $P^i_{\mu\nu}$, $i=1,2$ such that
\beq
P^i_{\mu\nu}W_A^{\mu\nu}=g_i  \;\;\;\; (i=1,2).
\eeq
One possible choice is the following:
\beqn
&&P^1_{\mu\nu}=P^{-1}\epsilon_{\mu\nu\alpha\beta}\,q^\alpha\,
\left(p^\beta+\frac{qs}{s^2}\frac{p^2}{pq}s^\beta\right)
\\
&&P^2_{\mu\nu}=P^{-1}\epsilon_{\mu\nu\alpha\beta}\,p^\alpha\,
\left(q^\beta-\frac{qs}{s^2}s^\beta\right),
\eeqn
with
\beq
P=2\sqrt{\abs{p^2}}qs\left[1-\frac{p^2q^2}{(pq)^2}
\left(1-\frac{(qs)^2}{q^2s^2}\right)\right].
\eeq
Notice that $P=0$ even for $p^2\ne 0$ if $s$
is purely longitudinal, as one can see using eqs.~(\ref{spin}) and
(\ref{spinnorm}); we must assume that $s$ has a transverse component
until we have projected out the structure functions, which are
by construction independent of $s$.
Collecting everything together, we finally obtain
\beqn
&&g_1(x,Q^2)=-P^{-1}\epsilon^{\mu\nu\gamma\delta}\,q_\gamma\,
\left(p_\delta+\frac{qs}{s^2}\frac{p^2}{pq}s_\delta\right)
\frac{1}{2\sqrt{\abs{p^2}}}\,
    \epsilon^{\rho\sigma\alpha\beta}p_\alpha s_\beta
W_{\mu\nu\rho\sigma}(p,q)
\label{g1}
\\
&&g_2(x,Q^2)=-P^{-1}\epsilon^{\mu\nu\gamma\delta}\,p_\gamma\,
\left(q_\delta-\frac{qs}{s^2}s_\delta\right)
\frac{1}{2\sqrt{\abs{p^2}}}\,
    \epsilon^{\rho\sigma\alpha\beta}p_\alpha s_\beta
W_{\mu\nu\rho\sigma}(p,q).
\label{g2}
\eeqn
Even before explicitly computing $W_{\mu\nu\rho\sigma}$, we can check
that $g_1$ and $g_2$ given by eqs.~(\ref{g1}) and (\ref{g2}) 
are indeed independent
of $s$, as they should. In fact, it is easy
to prove that only a term proportional to $g_{\mu\rho}p_\nu q_\sigma$
gives a non-zero contribution to $g_1$, while in the case of $g_2$
the only surviving term is proportional to $g_{\mu\rho}q_\nu q_\sigma$. 
Inserting these terms in eqs.~(\ref{g1}) and (\ref{g2}), the $s$-dependence
is seen to cancel against the factor $P^{-1}$.

\begin{figure}
\centerline{\epsfig{figure=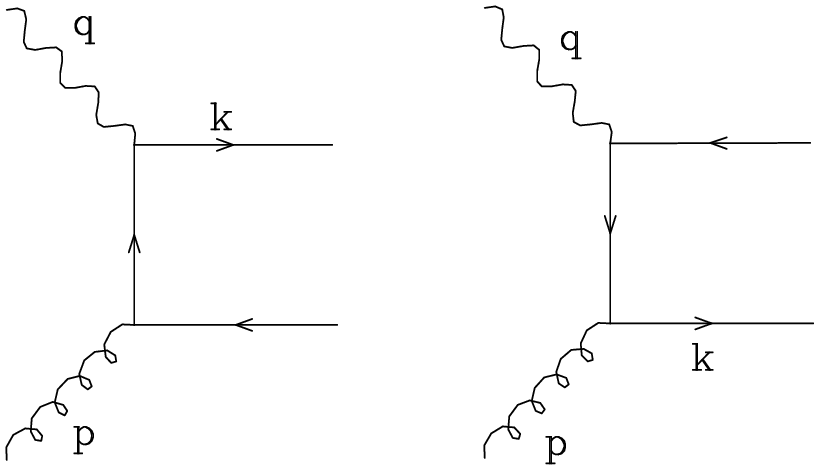,width=0.7\textwidth,clip=}}
\ccaption{}{ \label{fig1}  
Diagrams contributing to $W^{\mu\nu}$ for a gluon target.
}
\end{figure}

We now proceed to compute $W_{\mu\nu\rho\sigma}(p,q)$. The calculation
in this case is much simpler than in the quark case, because at order $\as$
there are no loop diagrams that contribute to the relevant amplitude.
For this reason, no ultraviolet or soft divergences are involved.
We have
\beq
\label{explw}
W_{\mu\nu\rho\sigma}=\frac{1}{4\pi}\frac{1}{N_c^2-1}
\sum_{colour} \int d\phi^{(2)} \left(A^{(1)}_{\mu\rho}+A^{(2)}_{\mu\rho}\right)
 \left(A^{(1)}_{\nu\sigma}+A^{(2)}_{\nu\sigma}\right)^*,
\eeq
where $N_c=3$ is the number of colours, and $d\phi^{(2)}$
is the two-body phase space. The amplitudes $A^{(1)}, A^{(2)}$ 
correspond to the diagrams of fig.~\ref{fig1}.
The singularities that arise when an on-shell gluon radiates massless quarks
in the collinear configuration may be regularized either by a
non-zero gluon virtuality $p^2$, or by a non-vanishing quark mass $m$.
We will keep both $p^2$ and $m$ different from zero at this level,
and we will discuss later the behaviour of our results in the limit 
$p^2,m^2\to 0$.
In the photon-gluon centre-of-mass frame, the momentum $k$ of the
produced quark is
\beq
k=\frac{E}{2}(1,0,\b\sin\theta,\b\cos\theta),
\eeq
where
\beq
E^2=(p+q)^2=p^2+Q^2\frac{1-x}{x}
\eeq
and
\beq
\b=\sqrt{1-\frac{4m^2}{E^2}}.
\eeq
The two-body phase space $d\phi^{(2)}$ takes the form
\beq
d\phi^{(2)}=\frac{\b}{16\pi}d\cos\theta.
\eeq
The denominators of the virtual quark propagators appearing in the
amplitude are given by
\beqn
\label{denq}
&&(k-q)^2-m^2=-pq\,\left(1+\b\bt\cos\theta\right)
\\
\label{denp}
&&(k-p)^2-m^2=-pq\,\left(1-\b\bt\cos\theta\right).
\eeqn
The phase-space integration is therefore singular
for $\cos\theta\to\pm 1$ when $\b=\bt=1$, or equivalently
$p^2=m^2=0$. The calculation is
straightforward (we have performed it with the help of the algebraic
manipulation program MACSYMA); the $\cos\theta$ integration can easily be
performed by observing that, after inserting eq.~(\ref{explw}) in 
eqs.~(\ref{g1}) and (\ref{g2}), the numerator of the integrand expression
is a degree-2 polynomial in the invariants $kq$, $kp$ and $ks$.
Terms proportional to powers of $kq$ and $kp$ can be integrated
immediately, since their dependence on $\cos\theta$ is given explicitly by
eqs.~(\ref{denq}) and (\ref{denp}). Terms containing
powers of $ks$ can also be
expressed in terms of integrals containing only $kq$ and $kp$. Consider
for example 
\beq
I^\mu=\int d\phi^{(2)}\,f(kp,kq)\,k^\mu,
\eeq
where $f(kp,kq)$ is a generic scalar function. 
The result must be a linear combination of $q^\mu$ and $p^\mu$:
\beq
I^\mu=A\,q^\mu+B\,p^\mu,
\eeq
and the scalar coefficients $A$ and $B$ can be obtained by solving the
system
\beqn
&&A\,q^2+B\,pq=q_\mu I^\mu
\\
&&A\,pq+B\,p^2=p_\mu I^\mu,
\eeqn
so that, finally,
\beq
s_\mu I^\mu=\int d\phi^{(2)}\,f(kp,kq)\,ks=A\,qs.
\eeq
Terms proportional to $(ks)^2$ can be treated in a similar way.
Therefore, all phase-space integrals are of the type
\beq
\int_{-1}^1 d\cos\theta\,
(1+\b\bt\cos\theta)^a(1-\b\bt\cos\theta)^b,
\eeq
with $a$ and $b$ integers between $-2$ and $2$. We obtain the
following results:
\beqn
g_1&=&-\frac{e^2 \as}{8\pi}\frac{1}{\bt^4 x}
\Bigg[\frac{\b}{1-\b^2 \bt^2} \,
(4 \bt^4 x^2-8 \b^2 \bt^2 x^2-8 \bt^2 x^2+12 x^2-2 \b^2 \bt^4 x
\nonumber\\
&&-6 \bt^4 x+8 \b^2 \bt^2 x+12 \bt^2 x-12 x-\bt^6+2 \b^2 \bt^4
+3\bt^4-2 \b^2 \bt^2-5 \bt^2+3)
\nonumber\\
&&-\frac{L}{2\bt} (4 \bt^4 x^2-8 \bt^2 x^2+12 x^2-4 \bt^4 x
+12 \bt^2 x-12 x-\bt^6+3 \bt^4-5 \bt^2+3)\Bigg]
\nonumber\\
\label{g1pm}
\\
g_2&=&-\frac{e^2 \as}{8\pi}\frac{1}{\bt^4 x}
\Bigg[\frac{\b}{1-\b^2\bt^2}\,
(8\b^2\bt^2 x^2+4\bt^2 x^2-12 x^2+2\b^2\bt^4 x
\nonumber\\
&&+2\bt^4 x-8\b^2\bt^2 x-8\bt^2 x+12x-2\b^2\bt^4-\bt^4+2\b^2\bt^2+4\bt^2-3)
\nonumber\\
&&-\frac{L}{2\bt} (4\bt^2 x^2-12 x^2-8\bt^2 x+12 x-\bt^4+4\bt^2-3)\Bigg],
\label{g2pm}
\eeqn
where $e$ is the electric charge of the produced quark in units of the
positron charge, $\as$ is the strong coupling, and
\beq
L=\log\frac{1+\b\bt}{1-\b\bt}.
\eeq
The collinear singularities are collected in the factor $L$, which
diverges logarithmically when both $m^2$ and $p^2$ go to zero.
The structure function $g_1$ was first computed in ref.~\cite{Kodaira}
for $m=0$, $p^2<0$. The general case $m^2\ne 0, p^2\ne 0$
was considered in ref.~\cite{Vogelsang}.
Our result for $g_1$, eq.~(\ref{g1pm}), is different from
the analogous formula obtained in ref.~\cite{Vogelsang}. The origin
of this discrepancy is
the fact that the operator used in ref.~\cite{Vogelsang} to obtain
$g_1$ from $W^{\mu\nu}_A$ actually projects out the desired structure
function only in the limit $p^2\to 0$, while $p^2$ is kept non-zero
elsewhere at this stage. However, the final result of ref.~\cite{Vogelsang} 
is correct in the physically interesting limit, as we shall see later.
Equation~(\ref{g2pm}), on the other hand, is a new result.

For $Q^2\to\infty$ with
\beq
r=\frac{-p^2}{m^2}
\eeq
fixed, we find
\beqn
g_1&=&-\frac{e^2 \as}{4\pi}\left[
\frac{4rx^3-6rx^2+2rx-4x+3}{rx^2-rx-1}-(2x-1)L\right]
\label{g1ex}
\\
g_2&=&+\frac{e^2 \as}{4\pi}\left[
\frac{6rx^3-10rx^2+4rx-4x+3}{rx^2-rx-1}-(2x-1)L\right].
\label{g2ex}
\eeqn
In this limit, $L$ takes the form
\beq
L=\log\frac{Q^2}{m^2}-\log\left(rx^2+\frac{x}{1-x}\right).
\eeq
It is interesting to notice that the terms proportional to $L$, which contains
the collinear divergence, cancel in the sum $g_1+g_2$.

As a test of the correctness of our calculation, we can check
that we reproduce the known results for $g_1$. Indeed, eq.~(\ref{g1ex}) gives
\beqn
&&r=0:\;\;\;
g_1=\frac{e^2 \as}{4\pi}\left[-4x+3+(2x-1)
\left(\log\frac{Q^2}{m^2}-\log\frac{x}{1-x}\right)\right],
\\
&&\phantom{r=0:\;\;\;}
\int_0^1 dx\, g_1(x,Q^2)=0;
\\
&&r\to\infty:\;\;\;
g_1=\frac{e^2 \as}{4\pi}\left[-4x+2+(2x-1)
\left(\log\frac{Q^2}{-p^2}-\log x^2\right)\right],
\\
&&\phantom{r=0:\;\;\;}
\int_0^1 dx\, g_1(x,Q^2)=-\frac{e^2 \as}{4\pi},
\eeqn
which coincide, for example, with the results of ref.~\cite{Vogelsang}.
The fact that different choices of the regularization
scheme lead to different results for $g_1$ (and in particular for
its first moment) has important physical
implications, and has been widely discussed in the
literature~\cite{g1}.

We now turn to the structure function $g_2$. From eq.~({\ref{g2ex})
we get
\beqn
&&r=0:\;\;\;
g_2=\frac{e^2 \as}{4\pi}\left[4x-3-(2x-1)
\left(\log\frac{Q^2}{m^2}-\log\frac{x}{1-x}\right)\right],
\\
&&r\to\infty:\;\;\;
g_2=\frac{e^2 \as}{4\pi}\left[6x-4-(2x-1)
\left(\log\frac{Q^2}{-p^2}-\log x^2\right)\right].
\eeqn
The first moment of $g_2$ vanishes in both cases; actually,
it is easy to prove, using eq.~(\ref{g2ex}), that
\beq
\int_0^1dx\, g_2(x,Q^2)=0
\eeq
for all values of $r$. Therefore we conclude that the BC
sum rule is satisfied also by the gluon contribution to $g_2$ at order $\as$,
within the class of regularization schemes we have adopted.

Having computed $g_2(x,Q^2)$, its $n$-th moment
\beq 
g_2^n=\int_0^1 dx\,x^{n-1}\,g_2(x,Q^2),
\eeq
can be obtained for any $n$.
In the two cases $r=0$ and $r\to\infty$, the $n^{th}$ moment of $g_2$
is given by
\beqn
&&r=0:\;\;\;
g_2^n=\frac{e^2 \as}{4\pi}
\left[\frac{4}{n+1}-\frac{3}{n}+\frac{1}{n^2}
-\frac{n-1}{n(n+1)}\left(\log\frac{Q^2}{m^2}-S(n)\right)\right]
\label{g2nzero}
\\
&&r\to\infty:\;\;\;
\label{g2ninf}
g_2^n=\frac{e^2 \as}{4\pi}
\left[\frac{6}{n+1}-\frac{4}{n}-\frac{4}{(n+1)^2}+\frac{2}{n^2}
-\frac{n-1}{n(n+1)}\log\frac{Q^2}{-p^2}\right],
\eeqn
where $S(n)=\sum_{k=1}^n 1/k$. 
Once again, we see that both expressions vanish when $n=1$.

The computation presented here is equivalent to the calculation
of gluon coefficient functions in the light-cone operator product expansion of
$W_A^{\mu\nu}$; in that case, the quantities that are directly 
computed are odd moments of the coefficient functions; in the case of
$g_2$, only moments for $n\geq 3$ are obtained, and therefore no direct
test of the BC sum rule can be performed.
Such a calculation was performed in ref.~\cite{Kodaira} in the case $m=0$,
which corresponds to our eq.~(\ref{g2ninf}). The two results are in agreement.

In conclusion, we have performed a calculation of the structure function
$g_2$ for a target gluon. We have considered various regularization schemes
for the collinear divergences, and we have found that in all of them the first
moment of $g_2$ vanishes as expected.

\section*{Acknowledgements}
We thank G.~Altarelli, M.~Anselmino, C.~Becchi, S.~Forte and S.~Frixione 
for interesting discussions.

\vfill\eject

\end{document}